\documentclass[showpacs,amssymb,twocolumn,floatfix,pre,aps,notitlepage]{revtex4-1}

\usepackage{amssymb,amsfonts,amsmath,bm}
\usepackage{graphics,graphicx}
\usepackage{xcolor}

\begin{document}

\title{A strong-coupling effective-field theory for asymmetrically charged plates with counterions only}

\author{Ladislav \v{S}amaj$^{1,2}$} 
\author{Emmanuel Trizac$^{2,3}$} 
\author{Martin Trulsson$^4$}

\affiliation{
$^1$Institute of Physics, Slovak Academy of Sciences, D\'ubravsk\'a cesta 9,
84511 Bratislava, Slovakia \\
$^2$LPTMS, Universit\'e Paris-Saclay, CNRS, 91405, Orsay, France \\
$^3$ENS de Lyon, 46 all\'{e}e d'Italie, 69364 Lyon, France \\ 
$^4$Computational Chemistry, Lund University, SE-221 00 Lund, Sweden}

\date{\today} 

\begin{abstract}
 {We are interested in rationalizing the phenomenon of like-charge attraction between charged bodies, such as a pair of colloids, in the strong coupling regime. The two colloids are modelled as uniformly charged parallel plates, neutralized by mobile counterions.  
In an earlier work [Palaia et al., J. Phys. Chem. B \textbf{126}, 3143 (2022)], we developed an effective-field theory for symmetric plates, stemming from the ground-state description that holds at infinite couplings. Here, we generalize the approach to the asymmetric case, where the plates bear charges of the same sign, but of different values.  
In the symmetric situation, the mobile ions, which are localized in the vicinity of the two plates, share equally between both of them. Here, the sharing is non-trivial, depending both on the coupling parameter and the distance between the plates. We thus introduce a counterion occupation parameter, that is determined variationally to ensure minimum
of the free energy. The resulting 
analytical results for the pressure as a function of the plate-plate distance $d$ agree well with
our Monte Carlo data, in a large interval of strong and intermediate
coupling constants $\Xi$.
We show in particular that within this description, there exists a range of large distances at which the attractive pressure features a $1/d^2$ behavior.
}
\end{abstract}

\maketitle

\renewcommand{\theequation}{1.\arabic{equation}}
\setcounter{equation}{0}

\section{Introduction} \label{Sec1}
Mesoscopic bodies (macroions or colloids), immersed in a polar solvent like
water, release from their surfaces (due to efficient solvation)
mobile ``counterions''.
Ions in Coulomb fluids are generically of both signs, however, one can reach
experimentally the limit of deionized (salt-free) suspensions
with no ``coions'' \cite{Raspaud00,Palberg04,Brunner04}.
The curved surface of the large colloid is usually approximated by
a planar surface and the modulated charge density fixed on colloid's
surfaces by the uniform one.
Counterions in the vicinity of a charged colloid form an electric
double layer (EDL) \cite{Attard96,Levin02,Messina09}.
The study of the effective interaction between two like-charged EDLs,
mediated by counterions, is of special experimental and theoretical interest
in many branches of physics, chemistry and biochemistry
\cite{BenTal95,Jonsson99,Hansen00,Belloni00,Grosberg02}.

Like-charged macroions always repel one another in the high-temperature
(weak coupling, WC) regime described by the mean-field Poisson-Boltzmann
(PB) theory \cite{Gouy10,Chapman13,rque51,Chan,Derjaguin87} as well as
its functional improvement via a loop expansion
\cite{Attard88,Podgornik90,Netz00}.

At  {low} enough temperatures, {\em i.e.,} in
the strong-coupling (SC) regime, a counter-intuitive attraction of
like-charged macromolecules was observed
by computer simulations
\cite{Guldbrand84,Kjellander84,Gronbech97} as well as
experimentally
\cite{Khan85,Kjellander88,Bloomfield91,Kekicheff93,Dubois98}.
Different theoretical treatments have been proposed for the SC regime.
In the virial SC approaches \cite{Moreira00,Moreira01,Netz01}, 
the leading SC term of the counterion density corresponds to a single 
particle theory in the electric potential of charged wall(s); 
resulting densities have been confirmed by Monte Carlo (MC) simulations
\cite{Kanduc07,Dean09,Moreira00,Moreira01,Naji05}.
Next correction orders in inverse powers of the coupling constant, obtained 
within a virial fugacity expansion, require a renormalization
of infrared divergencies; comparison with MC simulations shows that
the first correction term has the correct functional form in space,
but an incorrect prefactor.
Another type of SC theories was based on the classical Wigner crystal of
counterions created on the wall surfaces at zero temperature 
\cite{Grosberg02,Levin99,Shklovskii99}.
A harmonic analysis of counterion deviations from their ground-state Wigner
positions \cite{Samaj11a,Samaj11b} reproduces correctly the leading
single-particle picture of the virial SC approach.
The first correction term to the counterion density is in excellent agreement 
with MC data for strong as well as intermediate Coulombic couplings.
To adapt the Wigner SC approach to the fluid phase, the Wigner structure
was substituted by a correlation hole (i.e., the depletion region around
a charge due to Coulomb repulsion of the same charges) in
Refs. \cite{Samaj16,Palaia18}.

For two parallel symmetrically charged planar surfaces, it was recently shown
\cite{Palaia22} that the relevant physics for the like-charge attraction
is the ground state one.
 {The method \cite{Palaia22} is based on the introduction
of effective fields which reflect the partial screening of the electric field
induced by the fixed surface charge density of a plate by counterion layers.}   
According to Earnshaw's theorem \cite{Earnshaw}, in the ground state
counterions stick to the surfaces of the confining plates.
Upon changing the distance between plates from 0 to $\infty$,
a sequence of five Wigner phases I--V emerges at zero temperature
\cite{Goldoni96,Weis01,Messina03,Lobaskin07,Oguz07,Samaj12}.
These staggered phases consist of two equivalent lattice structures on
the left and right plates, shifted with respect to one another by
a half period in both spatial directions.
Since each plate as a whole (i.e., the surface charge density plus
the corresponding counterions) is electroneutral, the effective interaction
between the walls is short-ranged (exponentially decaying) at large distances.
The extension of the ground-state effective fields to nonzero
temperatures leads to a formula for the pressure which interpolates between
the ``ideal gas'' regime for small inter-plate distances and
the ``Wigner'' regime at large distances. 
The pressure fulfills known exact requirements and its dependence
on the inter-plate distance is in a perfect agreement with MC data, in
a large interval of strong and intermediate values of the coupling constant.

The aim of this paper is to extend the effective-field method \cite{Palaia22}
to asymmetric parallel plates;  {throughout this paper,
asymmetrically charged plates refers to surfaces with unequal but same-sign surface 
charge densities}.
The ground state of asymmetric plates was studied by using analytical and
computational evolutionary techniques \cite{Antlanger16,Antlanger18}, as
well as unsupervised learning \cite{Hartl23}. 
In comparison with the symmetrically charged plates, the asymmetric system
exhibits much more phases, sometimes of exotic nature (pentagonal, snub square etc.). 
Each plate as a whole  {(i.e., the surface charge of the plate
plus the counterions attached to that plate)} is, in general, not neutral
which implies a long-ranged (inverse-power law) effective interaction between
the plates at large distances between them.
This non-neutrality phenomenon complicates substantially
the analytic treatment of the asymmetric problem because of the presence
of an additional free parameter into the theory, namely
the one related to the counterion occupations of the plates.
In the ground state, this parameter is determined variationally
to ensure minimum of the ground-state energy.
In this paper, we go to nonzero-temperature and construct
the free energy of the system in the strong coupling regime,
in terms of deviations of counterions from their ground-state positions.
The  {counterion} occupation parameter is determined
variationally to ensure minimum of the free energy.
Applying then the effective-field idea \cite{Palaia22}, analytic results
for the pressure as a function of the distance $d$ between the plates
agree very well with our MC data in a large interval of strong and
intermediate coupling constants.
For large enough distances $d$ of the {\em attractive} regime, the pressure
is shown to scale like $1/d^2$, with the non-universal prefactor which carries
the structural information about the Wigner ground state and depends
on the asymmetry parameter of the plates.
We stress that at asymptotically large distances, the pressure is expected to
follow the Poisson-Boltzmann behavior, and to be repulsive there,
decaying as $1/d^2$.

The paper is organized as follows.
Sec.~\ref{Sec2} brings basic  {setup} for the asymmetric model,
together with the notation used.
The ground-state structures for symmetrically and asymmetrically charged
plates are summarized in Sec.~\ref{Sec3}.
The emphasis is put on the regions of small and large inter-plate distances
characterized by a few notable bilayer phases.
 {Although sections \ref{Sec2} and \ref{Sec3} summarize
in a relatively detailed way the known results from previous papers,
they make the presentation self-contained.
The presented formulas are crucial for a clear understanding of original
results derived in the succeeding sections and help the potential reader
to reproduce the obtained analytic results.}
Sec.~\ref{Sec4} concerns the analytic effective-field treatment of
the asymmetric model at nonzero temperatures, within the SC regime.
The pressure is obtained from either the contact value theorem
(Sec.~\ref{Sec41}) or the thermodynamic route (Sec.~\ref{Sec42}).
Details of MC simulations performed in this paper are described
in Sec.~\ref{Sec5}.
The comparison of the analytic and MC results for intermediate values
of the coupling constant $\Xi = 30, 100$ is made in Sec.~\ref{Sec6}.
The emphasis is put on both the small-distance region, where
the attractive pressure exhibits its minimum, as well as large-distance
region, where the attractive pressure exhibits the $1/d^2$ decay
with a non-universal prefactor.
Sec.~\ref{Sec7} is a short recapitulation, together with some concluding
remarks.

\renewcommand{\theequation}{2.\arabic{equation}}
\setcounter{equation}{0}

\begin{figure}[tbp]
\begin{center}
\includegraphics[clip,width=0.48\textwidth]{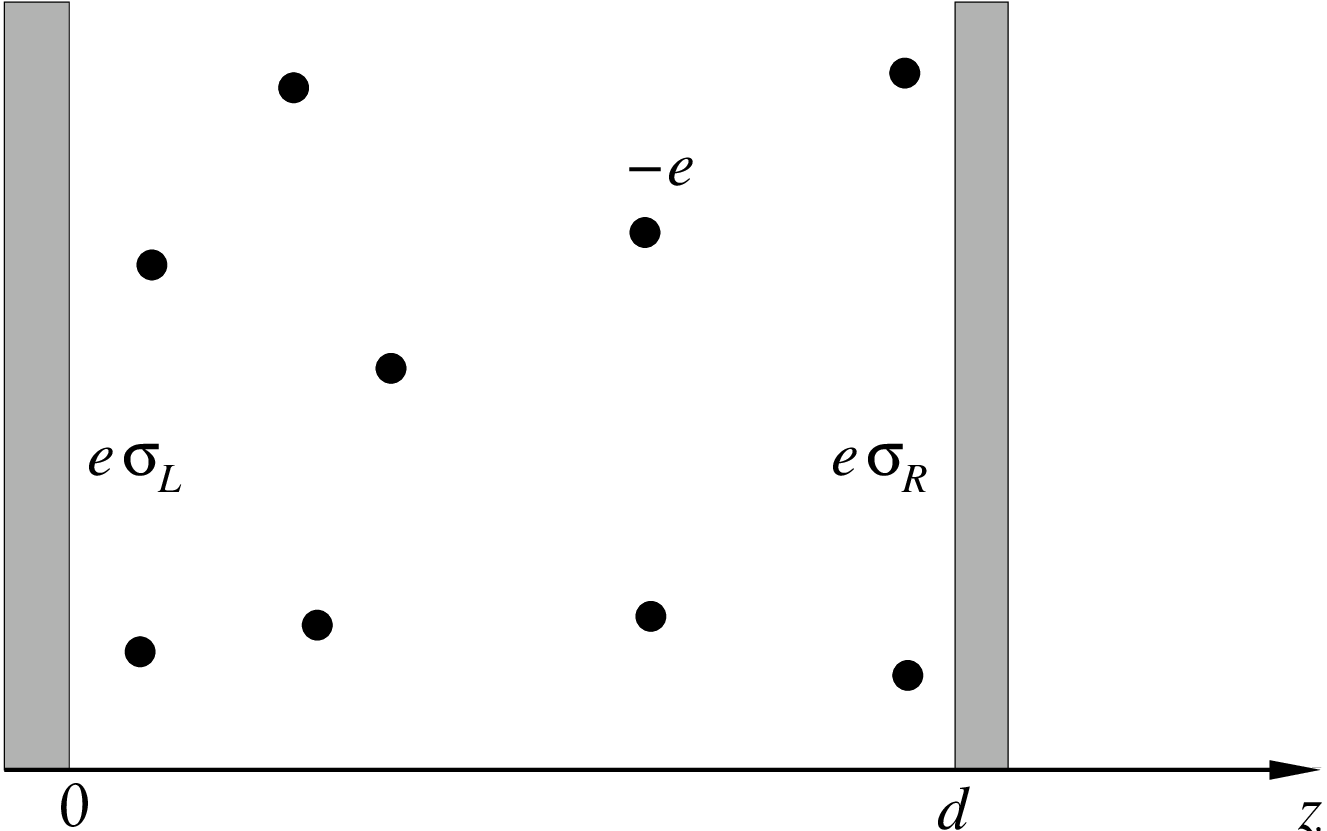}
\caption{The geometry (along the $z$-axis) of two parallel walls
at distance $d$.
There is a homogeneous surface charge density $e \sigma_L$ fixed on
the left wall and $e \sigma_R$ on the right wall.
The pointlike counterions of charge $-e$, moving freely between the walls,
are pictured as black circles.}
\label{fig1}
\end{center}
\end{figure}

\section{Basic setup for the asymmetric model} \label{Sec2}

\subsection{Notation}
Let us consider a pair of parallel plates at distance $d$, in the 3D
Cartesian space of points ${\bf r}=(x,y,z)$, see Fig. \ref{fig1}.
The left and right plates of the same (large) surface $S$ spread along
the 2D plane $(x,y)$, their positions along the perpendicular $z$-axis
being $0$ and $d$, respectively.
The left (right) plate carries a homogeneous surface charge density
$e \sigma_L$ ($e \sigma_R$) where $e$ is the elementary charge.
 {
Colloids acquire, in most cases, their charge from a chemical equilibrium, of \emph{e.g.} ions or charged molecules, between their surface and the solution medium; this equilibrium depends, in general, on the thermodynamic parameters \cite{Chan,Derjaguin87}.
The investigation of the surface charge regulation, where $\sigma_R$ and $\sigma_L$ would no longer be constant, goes beyond the scope of
this work.
We note that some surfaces like mica are structural and cannot change
their charge density; this includes a number of mineral surfaces.
For titratable surfaces the surface charge would indeed vary, but this relies
also on a salt reservoir which we do not allow for.
Nevertheless, in the most frequent common chemical equilibrium
when the solution's pH is far away from the colloid's pKa-values,
the colloid can essentially be regarded as having a fixed surface charge.}

There are $N$ classical counterions with charge say of unit valence $-e$
which move in the space between the plates $\Lambda = \{ {\bf r},0<z<d \}$.
The requirement of the overall neutrality reads
\begin{equation} \label{neutrality}
N = (\sigma_L + \sigma_R) S .
\end{equation}
 {The counterions are considered to be pointlike.
This simplification is suitable especially for low temperatures when
counterions maximize their separation within the counterion layer and
their size will be irrelevant up to the spacing of the counterion structure
\cite{Samaj20}.}

Without any loss of generality one can assume that $\sigma_L>0$.
Rescaling appropriately the model's parameters, it is sufficient
to consider the asymmetry parameter
\begin{equation}
A = \frac{\sigma_R}{\sigma_L}
\end{equation}  
inside the interval $[-1,1]$; the limiting value $A=-1$ corresponds to
the trivial case $\sigma_R=-\sigma_L$ with no counterions between the plates,
$A=1$ corresponds to the symmetrically charged plates $\sigma_L=\sigma_R$.
Likely charged asymmetric plates with
\begin{equation}
0 < \sigma_R < \sigma_L  ,
\end{equation}  
i.e. $A\in (0,1)$, are of special interest.
Let the dielectric constant of the walls $\varepsilon_W$ be the same as that
of the medium the counterions are immersed in $\varepsilon$,
$\varepsilon_W=\varepsilon$, i.e., there are no image charges. 
The vacuum $\varepsilon=1$ is taken for simplicity, again without loss
of generality. 
The system is considered either at zero temperature $T=0$ or
in thermal equilibrium at nonzero temperatures $T>0$.

\subsection{Zero temperature, $T=0$}
In the ground state ($T=0$), according to Earnshaw's theorem \cite{Earnshaw}, 
the Coulomb charges are expelled from the slab interior and stick to
the surfaces of the confining plates.
In particular, $N_L$ ($N_R$) counterions collapse on the left (right) plate
surfaces, $N = N_L + N_R$ and organize themselves onto certain left (right)
crystal structures.
The densities of counterions at the surfaces of the plates are given by
\begin{equation}
n_L = \frac{N_L}{S} , \qquad n_R = \frac{N_R}{S} . 
\end{equation}  
According to the overall electroneutrality condition (\ref{neutrality}),
the  {counterion} densities at plates are constrained by
\begin{equation}
n_L + n_R = \sigma_L + \sigma_R .
\end{equation}
It is useful to introduce  {the (occupation) order parameter}
\begin{equation} \label{x}
p \equiv \frac{N_R}{N_L+N_R} = \frac{n_R}{\sigma_L+\sigma_R}
\end{equation}
by using of which one can express $n_L$ and $n_R$ as follows
\begin{eqnarray}
n_L & = & (1-p) (\sigma_L + \sigma_R) \label{nL}  \\
n_R & = & p (\sigma_L + \sigma_R) . \label{nR} 
\end{eqnarray}
If the local electroneutrality holds on both plates, i.e.,
$n_L=\sigma_L$ and $n_R=\sigma_R$, then the  {order}
parameter $p$ equals to 
\begin{equation}
p_{\rm neutr} = \frac{A}{1+A} .  
\end{equation}

\begin{itemize}
\item  
For the symmetric case $A=1$ with $\sigma_L=\sigma_R=\sigma$,
upon changing the distance between plates from 0 to $\infty$,
a sequence of five phases I--V emerges
\cite{Goldoni96,Weis01,Messina03,Lobaskin07,Oguz07,Samaj12}.
These staggered phases consist of two equivalent lattice structures on
the left and right plates, shifted with respect to one another by
a half period in both spatial directions.
Phase I, the monolayer hexagonal structure, exists only at zero
inter-plate distance $d=0$ \cite{Messina03,Samaj12}.
Phase II corresponds to a staggered rectangular bilayer with
the aspect ratio $1<\Delta<\sqrt{3}$, phase III is a staggered
square bilayer, phase IV a staggered rhombic bilayer with a deformation
angle $\varphi<\pi/2$ and phase V a staggered hexagonal bilayer. 
Since $N_L=N_R=N/2$ for each of the phases, it holds that $n_L=n_R=\sigma$
or, equivalently, $p_{\rm gs}=\frac{1}{2}$ where the subscript ``gs'' means
``ground-state''.
Each plate as a whole (i.e., the surface charge density plus the counterions)
is thus neutral.
As a consequence, the walls are bounded at large distances by short-ranged
(usually exponentially decaying) forces.
\item
The asymmetric case $0<A<1$, studied by using both analytical calculations
and computational evolutionary techniques as well as unsupervised learning
in Refs. \cite{Antlanger16,Antlanger18,Hartl23}, exhibits much more phases. 
In an interval of small distances up to a critical one, $0\le d<d_c(A)$,
phase I with all counterions collapsed onto the hexagonal monolayer on
the left plate is dominant, i.e.,
\begin{equation} \label{limitzero}
n_L = \sigma_L + \sigma_R , \qquad n_R = 0 .
\end{equation}
Consequently,
\begin{equation} \label{xgs}
p_{\rm gs}(d) = 0   \qquad \mbox{for $0\le d \le d_c(A)$,}
\end{equation}
In the opposite asymptotic limit of large distances $d\to\infty$, the local
neutralization of each of the plates  {by the corresponding
counterions} takes place: 
\begin{equation} \label{limitinfty}
\lim_{d\to\infty} n_L(d) = \sigma_L , \qquad
\lim_{d\to\infty} n_R(d) = \sigma_R .
\end{equation}  
Consequently,
\begin{equation} \label{smallasymx}
\lim_{d\to\infty} p_{\rm gs}(d) = p_{\rm neutr} .
\end{equation}
As a rule, $p_{\rm gs}(d)$ grows from 0 to $p_{\rm neutr}$ monotonously
with increasing distance $d$.
In general, for finite inter-plate distances $d$, the counterion densities
$n_L\ne \sigma_L$ and $n_R\ne \sigma_R$ do not neutralize locally
the corresponding surface charge densities at the plates. 
This implies long-ranged (inverse-power law) effective interactions between
the plates at large distances \cite{Antlanger16,Antlanger18}.
\end{itemize}

\subsection{Nonzero temperatures, $T>0$}
As soon as $T>0$, the counterions can move into the slab interior.
When the temperature is  {low}, the counterions are still
localized in the neighborhood of the plate they belonged at zero temperature
and one can adopt plausibly the counterparts of the densities $n_L$ and $n_R$;
they do not represent the counterion densities {\em at} the walls
but rather {\em under the effect of the effective fields} generated by
the corresponding walls.
\begin{itemize}
\item  
For the symmetrically charged plates, the reflection symmetry of the system
keeps the local electroneutrality of the plates, $n_L=n_R=\sigma$,
and so, likewise in the ground state, $p=\frac{1}{2}$.
This is behind the success of the method of interest \cite{Palaia22} to
describe the thermodynamics for nonzero temperatures in terms of
ground-state effective fields. 
\item
For asymmetrically charged plates, in close analogy with the zero temperature,
the  {counterion} densities are not expected to neutralize
locally the corresponding surface charge densities at the plates for
finite distances, i.e., in general, $n_L\ne \sigma_L$ and $n_R\ne \sigma_R$. 
A crucial complication for asymmetric plates is a discontinuous change
of counterion densities $n_L$ and $n_R$ when passing from zero to
nonzero temperatures.
 {As will be shown later both analytically as well as numerically, the counterion densities skip from values $n_L=(\sigma_L+\sigma_R)$ and $n_R=0$ to $n_L=n_R = (\sigma_L+\sigma_R)/2$ at asymptotically small distances, as soon as $T$ goes from 0 to a nonzero value.}
This is because for small inter-plate distances the
counterions move freely in the electric field created by
the uniformly charged plates (the potential difference between the plates
is very small) and  {
 the neighbouring counterions are sufficiently far away from each other for having any effect on the electric field in the  direction perpendicular to the plates.}
Consequently,
\begin{equation} \label{xd}
p(d) \sim \frac{1}{2} \qquad \mbox{for $d\to 0^+$.}
\end{equation}
We conclude that for small distances going from $T=0$ to $T>0$ induces
a discontinuity in $p$ from $p_{\rm gs}=0$ (\ref{xgs}) to $p=\frac{1}{2}$
(\ref{xd}), respectively.
To extend the ground-state description in terms of effective fields to
nonzero temperatures is then nontrivial due to atypical entropy contributions.
\end{itemize}

\renewcommand{\theequation}{3.\arabic{equation}}
\setcounter{equation}{0}

\section{Ground-state picture} \label{Sec3}
To neutralize the plate surface with a fixed surface charge density $e \sigma$
by a regular lattice structure of point charges $-e$, the lattice constant
should be of order $1/\sqrt{\sigma}$.
At zero temperature, the distance between the plates will be considered in
the dimensionless form
\begin{equation} \label{eta}
\eta = d \sqrt{\frac{\sigma_L+\sigma_R}{2}} .
\end{equation}  
Roughly speaking, the parameter $\eta$ is the ratio of the distance between
the plates $d$ and the characteristic distance between the nearest neighbor
counterions on the plates.

 {There are
two ways to obtain the ground state pressure: either via the counterion density at the wall contact (contact theorem, see below), or the energy change with the inter-plate distance.}

\subsection{The contact pressure} \label{Sec31}
 {Under the term ``effective field'' we understand the factor
by which the electric field created by the fixed surface charge density of a plate
is screened by counterion layers.}   
Let us derive first, in Gaussian units, the effective field acting on
counterions constrained to the left plate.
The electric field generated by the uniform surface charge density $e\sigma_L$
is given by
\begin{equation} \label{EL}
E_L = 2\pi e\sigma_L . 
\end{equation}  
For a single counterion on the left plate ($z=0$), the layer of counterions on
the same plate induces a symmetric potential $V(z)=V(-z)$ and therefore
the electric field $E=-\partial V(z)/\partial z$, proportional to $z$,
is subdominant with respect to (\ref{EL}).
The discrete layer of ions on the opposite right plate, together with the
uniform surface charge density $e \sigma_R$ on that plate, renormalize the bare
field $E_L$ by a factor $\kappa_L$ which depends on the distance
$\eta$.
For small distances $\eta\to 0$, each counterions on the left plate feels
the electric field generated by the two plates only
$2\pi e(\sigma_L-\sigma_R)$, while discrete layers of other
counterions are too far away compared to the inter-plate distance to contribute to the energy, i.e.,
$\kappa_L=(\sigma_L-\sigma_R)/\sigma_L$.
For large distances $\eta\to\infty$, the discrete character of the
 {counterion} layer on the opposite right plate becomes
irrelevant and together with the fixed surface charge density they form
a neutral entity, i.e., $\kappa_L=1$.
To summarize,
\begin{equation} \label{limits1}
\kappa_L \mathop{\sim}_{\eta\to 0} 1 - A  ,
\qquad \kappa_L \mathop{\sim}_{\eta\to \infty} 1 .  
\end{equation}
Note that these limiting values of the effective field are not restricted
to the ground state, but they apply also to nonzero temperatures.

Each ion at the contact with the left plate pushes on it with a force
$\kappa_L e E_L$; since there are $n_L$ ions per unit surface,
the repulsive force per unit surface is $\kappa_L e n_L E_L$. 
On the other hand, there is an electrostatic force acting on
the left plate due to the presence of two (left and right) ion layers
and of the surface charge on the right plate.
Since the corresponding surface charge density $- e \sigma_L$ is opposite
to the original one on the left plate, the attractive force per unit surface
is $- 2\pi (e \sigma_L)^2 = - e \sigma_L E_L$.
The total force per unit surface, i.e. the pressure, is the sum of
the contact and electrostatic forces:
\begin{eqnarray}
P_0 & = & \kappa_L e n_L E_L - e \sigma_L E_L
\nonumber \\ & = & 2\pi e^2 \left( \kappa_L \sigma_L n_L
- \sigma_L^2 \right) . \label{press11}
\end{eqnarray}  
In this paper, we follow the convention of Refs. \cite{Samaj11b,Kanduc08}
that all thermodynamic quantities will be rescaled to their dimensionless
forms with respect to the {\em left} plate.
In particular, the pressure will be considered in the dimensionless form
\begin{equation} \label{tildeP}
\widetilde{P}_0 \equiv \frac{P_0}{2\pi e^2\sigma_L^2} .  
\end{equation}
Thus the dimensionless form of the relation (\ref{press11}) reads as
\begin{equation} \label{press1}
\widetilde{P}_0 = \frac{\kappa_L n_L}{\sigma_L} - 1 .
\end{equation}

An analogous analysis for the counterions on the right plate implies that
\begin{equation} \label{press2}
\widetilde{P}_0 =  A^2 \left( \frac{\kappa_R n_R}{\sigma_R} - 1 \right) . 
\end{equation}
Here, the factor $\kappa_R$ renormalizes the bare field induced by
the right plate $E_R = -2\pi e\sigma_R$ due to the presence
of the homogeneous surface charge density and the discrete layer of
ions on the opposite left plate.
The counterparts of the limiting values (\ref{limits1}) read as 
\begin{equation} \label{limits2}
\kappa_R \mathop{\sim}_{\eta\to 0} 1 - \frac{1}{A} ,
\qquad \kappa_R \mathop{\sim}_{\eta\to \infty} 1 .  
\end{equation}  

The two equivalent relations for the pressure (\ref{press1}) and
(\ref{press2}) imply the equality
\begin{equation} \label{equal}
\kappa_L n_L\sigma_L - \kappa_R n_R \sigma_R
= \sigma_L^2 - \sigma_R^2   
\end{equation}
or, equivalently,
\begin{equation} \label{equalprime}
\kappa_L (1-p) - \kappa_R A p = 1-A .  
\end{equation}

\subsection{The pressure obtained via the thermodynamic route} \label{Sec32}
The previous definitions (\ref{press1}) and (\ref{press2}) of the dimensionless
pressure were given in terms of the quantities at the plate contacts.
Let us introduce the auxiliary quantity $E_0(\eta,p)$ as the ground-state
energy per unit surface in the subspace with a fixed  {order}
parameter $p$.
One can define the pressure alternatively as (minus) the total derivative of
this energy with respect to the distance:
\begin{eqnarray} 
P_0(\eta,p) & = & - \frac{\rm d}{{\rm d} d} \frac{E_0(\eta,p)}{S}
\nonumber \\ & = &
- \frac{e^2}{\sqrt{2}} (\sigma_L+\sigma_R)^2 \frac{\rm d}{{\rm d}\eta}
\frac{E_0(\eta,p)}{N e^2\sqrt{\sigma_L+\sigma_R}} ; \phantom{aaaa} \label{press}
\end{eqnarray}
hereinafter, if not necessary, the explicit dependence of quantities
on the asymmetry parameter $A$ will not be indicated.
The ground-state value of $p$ is determined by the condition of
the energy minimum:
\begin{equation} \label{energymin}
\frac{\partial E_0(\eta,p)}{\partial p}\Big\vert_{p=p_{\rm gs}} = 0 , \qquad
\frac{\partial^2 E_0(\eta,p)}{\partial p^2}\Big\vert_{p=p_{\rm gs}} < 0 .  
\end{equation}
The dimensionless pressure (\ref{tildeP}) is thus expressible as  
\begin{equation} \label{dimpress}
\widetilde{P}_0(\eta,p) = - \frac{1}{2^{3/2}\pi} (1+A)^2
\frac{\partial}{\partial \eta}
\frac{E_0(\eta,p)}{N e^2\sqrt{\sigma_L+\sigma_R}} ,
\end{equation}
where the interchange of the total derivative by the partial one is possible
due to the stationarity condition (\ref{energymin}). 
In the ground-state, the physical values of all considered quantities
is taken at $p=p_{\rm gs}$, in particular
\begin{equation}
\widetilde{P}_{\rm gs}(\eta) = \widetilde{P}_0(\eta,p_{\rm gs}) . 
\end{equation}
  
From (\ref{press1}) one can express the combination
\begin{equation}
\frac{\kappa_L n_L}{\sigma_L} = 1 + \widetilde{P}_0(\eta,p) .
\end{equation}
With regard to (\ref{nL}), it holds that
\begin{eqnarray}
\kappa_L(\eta,p) & = & \frac{1}{(1-p)(1+A)}
\left[ 1 + \widetilde{P}_0(\eta,p) \right] \nonumber \\
& = & 1 - A + A p K_0(\eta,p) , \label{kappaL}
\end{eqnarray}
where the function $K_0(\eta,p)$ is defined by
\begin{equation} \label{Kgs}
K_0(\eta,p) = \frac{\widetilde{P}_0(\eta,p) + A^2
+ p(1-A^2)}{p (1-p) A (1+A)} .  
\end{equation}
Similarly, it follows from (\ref{press2}) that
\begin{equation}
\frac{ \kappa_R n_R}{\sigma_R} =
1 + \frac{1}{A^2} \widetilde{P}_0(\eta,p) . 
\end{equation}
Since $n_R=p(\sigma_L+\sigma_R)$, one ends up with
\begin{eqnarray}
\kappa_R(\eta,p)  & = & \frac{1}{p(1+A)} \left[ A + \frac{1}{A}
\widetilde{P}_0(\eta,p) \right] \nonumber \\
& = & 1 - \frac{1}{A} + (1-p) K_0(\eta,p) . \label{kappaR}
\end{eqnarray}  
The ground-state values of the effective fields are given by
\begin{equation}
\kappa_L^{\rm (gs)}(\eta) = \kappa_L(\eta,p_{\rm gs}) , \qquad
\kappa_R^{\rm (gs)}(\eta) = \kappa_R(\eta,p_{\rm gs}) .  
\end{equation}

\subsection{Ground-state structures} \label{Sec33}
The above discussion was quite general, valid for any type of the bilayer.
In what follows, we shall restrict ourselves to the description of the
relevant phases present in the phase diagram at small and large values of
the distance $\eta$.
The exotic (snub square, pentagonal, ...) phases, taking place at intermediate
inter-plate distances, are irrelevant for our purposes.

According to the general analysis presented in Appendix A of
Ref. \cite{Antlanger18}, the total energy $E(\eta,p,A)$ of any bilayer system
with non-neutral (surface charge plus counterions) plates  can be expressed
in terms of the total energy of the bilayer system with ``neutralized''
plates, keeping the same values of $\eta, p$ and fixing the neutral value of
the asymmetry parameter $A=p/(1-p)$, as follows
\begin{eqnarray} 
\frac{E_0(\eta,p;A)}{N e^2 \sqrt{\sigma_L+\sigma_R}} & = &
\frac{E_0^{\rm neutr}\left(\eta,p;A=p/(1-p)\right)}{N e^2 \sqrt{\sigma_L+\sigma_R}}
\nonumber \\ & &
+ 2^{3/2} \pi \eta \left( p - \frac{A}{1+A} \right)^2 . \label{neutral}
\end{eqnarray}
The second term on the rhs of this equation is simply the excess
energy due to the non-neutrality of each of the two plate's entities.

\subsubsection{Structures {\rm I} and ${\rm I}_p$ emerging at small $\eta$}
As was mentioned in the Introduction, at small distances
$\eta\in [0,\eta_c(A)]$, all counterions collapse onto the left plate in
the so-called phase I \cite{Antlanger16,Antlanger18},
see the relations (\ref{limitzero}) implying that $p_{\rm gs}=0$. 
Here, $\eta_c(A)$ is a critical distance at which a second-order transition
from phase I to another one with $p_{\rm gs}>0$ and a smaller energy
takes place.
The value of $\eta_c(A)$ increases with decreasing $A$; it goes from
$\eta_c=0$ for the symmetrically charged $A=1$ plates up to
$\eta_c\to\infty$ when $A=0$ ($\sigma_R=0$).
The lattice spacing $a$ of the hexagonal structure of counterions at
the left plate is determined by the relation
$\sqrt{3} a^2 (\sigma_L+\sigma_R)/2 = 1$.
The energy of phase I, $E_{\rm I}(\eta,A)$, is given by
\cite{Antlanger16,Antlanger18}
\begin{equation}
\frac{E_{\rm I}(\eta,A)}{N e^2 \sqrt{\sigma_L+\sigma_R}} = c
+ 2^{3/2}\pi\eta \left( \frac{A}{1+A} \right)^2 
\end{equation}
with
\begin{eqnarray}
c & = & \frac{1}{2^{3/2}\sqrt{\pi}} \int_0^{\infty} \frac{{\rm d}t}{\sqrt{t}}
\left\{ \left[ \theta_3({\rm e}^{-\sqrt{3}t}) \theta_3({\rm e}^{-t/\sqrt{3}})
-1-\frac{\pi}{t} \right] \right. \nonumber \\ & &
+ \left. \left[ \theta_2({\rm e}^{-\sqrt{3}t}) \theta_2({\rm e}^{-t/\sqrt{3}})
-\frac{\pi}{t} \right] \right\} \nonumber \\
& = & -1.960515789\ldots  
\end{eqnarray}
being the Madelung constant of the 2D hexagonal structure;
here, $\theta_2(q) = \sum_{j=-\infty}^{\infty} q^{\left( j-\frac{1}{2} \right)^2}$
and $\theta_3(q) = \sum_{j=-\infty}^{\infty} q^{j^2}$ are Jacobi theta functions
with zero argument.
According to (\ref{dimpress}), the dimensionless pressure
\begin{equation}
\widetilde{P}_{\rm gs}(\eta) = - A^2
\end{equation}
is constant in phase I.
The same result follows directly by inserting $n_R=0$ into the formula
(\ref{press2}) which confirms the consistency of the ground-state formalism.

When $\eta$ exceeds its critical value $\eta_c(A)$, some of the counterions
on the left plate start to jump perpendicularly to the right plate,
i.e., the projections of the counterions of both layers onto
one plane still form the hexagonal lattice.
The corresponding phase with the given value of $p$ is referred to as
phase ${\rm I}_p$.
There are specific ``commensurate'' values of
$p\in \{ \frac{1}{2},\frac{1}{3},\frac{1}{4},\frac{1}{7},\frac{1}{9},...\}$
for which the counterions on the right place form an energetically
favorable hexagonal lattice with spacing $b\ge a$.
The interaction Coulomb energy of phase ${\rm I}_p$ is presented in
Appendix \ref{appA}.
Since the distribution of commensurate values of $p$ becomes denser
and denser as $p\to 0$, it is natural to extend the formulas for
the energy (\ref{appA1}) and (\ref{appA2}) to continuous values of $p$
in this limit of $p\to 0$.

In the region of small $p$, the expression for the energy (\ref{appA1})
can be expanded systematically in powers of $p$ \cite{Antlanger18}:
\begin{equation} \label{EIx}
\frac{E_{{\rm I}_p}(\eta,p)-E_{\rm I}(\eta,p)}{N e^2\sqrt{\sigma_L+\sigma_R}}
= f(\eta) p + \frac{2^{3/2}\pi}{\lambda} \eta^2 p^{5/2}
+ {\cal O}\left( p^{7/2}\right) ,
\end{equation}  
where
\begin{eqnarray}
f(\eta) =  2^{3/2}\pi \frac{1-A}{1+A} \eta - \frac{1}{\sqrt{2\pi}}
\int_0^{\infty} \frac{{\rm d}t}{\sqrt{t}} \left( 1 - {\rm e}^{-\eta^2 t} \right)
\phantom{aaaaaa} \nonumber \\ \times \left[
\theta_3({\rm e}^{-\sqrt{3}t}) \theta_3({\rm e}^{-t/\sqrt{3}}) -1
+ \theta_2({\rm e}^{-\sqrt{3}t}) \theta_2({\rm e}^{-t/\sqrt{3}}) \right]
\nonumber \\ \label{feta}
\end{eqnarray}
and
\begin{equation}
\lambda = \frac{4\pi}{3^{1/4}\zeta\left(\frac{3}{2}\right) \left[
\zeta\left(\frac{3}{2},\frac{1}{3}\right) -
\zeta\left(\frac{3}{2},\frac{2}{3}\right) \right]} \simeq 0.999215 .
\end{equation}
Here, $\zeta(s,q)=\sum_{j=0}^{\infty} 1/(q+j)^s$ is the Hurwitz zeta
function which represents the generalization of the Riemann zeta function
$\zeta(s)\equiv \zeta(s,1)$.
The extremum condition for the energy of phase ${\rm I}_p$ (\ref{EIx})
reads as
\begin{equation} \label{extremumcond}
0 = f(\eta) + \frac{5\sqrt{2}\pi}{\lambda} \eta^2 p^{3/2} +
{\cal O}\left( p^{5/2}\right) .  
\end{equation}  

For a given $A$, the critical value of the dimensionless distance
is identified with the condition $f(\eta_c)=0$, i.e.,
\begin{eqnarray}
4 \pi \frac{1-A}{1+A} \eta_c  =  \frac{1}{\sqrt{\pi}}
\int_0^{\infty} \frac{{\rm d}t}{\sqrt{t}} \left( 1 - {\rm e}^{-\eta_c^2 t} \right)
\phantom{aaaaaaaaaaaaa} \nonumber \\ \times
\left[ \theta_3({\rm e}^{-\sqrt{3}t}) \theta_3({\rm e}^{-t/\sqrt{3}}) -1
+ \theta_2({\rm e}^{-\sqrt{3}t}) \theta_2({\rm e}^{-t/\sqrt{3}}) \right] .
\nonumber \\ 
\end{eqnarray}
The function $f(\eta)$ (\ref{feta}) is dominated by the positive linear term
for small $\eta$, so that $f(\eta)>0$ for $0\le \eta<\eta_c$
and $f(\eta)<0$ for $\eta>\eta_c$.
In the region $0\le \eta<\eta_c$, in the vicinity of the critical point,
$f(\eta)$ can be expanded as $f(\eta) \sim g(\eta_c-\eta)$ with a positive
prefactor $g>0$.
In the region $0<\eta<\eta_c$, the extremum equation (\ref{extremumcond})
has no real solution for $p$; the minimum energy is determined by $p=0$
(phase I) in that region.
On the other hand, the extremum equation (\ref{extremumcond}) has
a positive (real) solution for $p$ in the region $\eta>\eta_c$,
$p(\eta)\propto (\eta-\eta_c)^{2/3}$, which grows continuously from 0
at $\eta=\eta_c$.
It is simple to verify that this solution provides the minimum of
the energy within the phase ${\rm I}_p$.

\subsubsection{Structure ${\rm V}_p$ emerging at large $\eta$}
The energy of phase ${\rm V}_p$ is presented in Eq. (\ref{EV}).
The large-$\eta$ asymptotic of the integral (\ref{J}) was calculated
by using the saddle-point method in Appendix E of Ref. \cite{Antlanger18},
with the result
\begin{equation}
J(\eta,p) \mathop{\sim}_{\eta\to\infty} - \frac{3^{5/4}}{\sqrt{2}} p \sqrt{1-p}
\exp\left( - \frac{4\pi\sqrt{1-p}}{3^{1/4}} \eta \right) .
\end{equation}
This integral decays exponentially in $\eta$ and therefore it can be neglected
comparing the term of the order $\eta$ in (\ref{EV}), i.e.,
\begin{eqnarray}
\frac{E_{{\rm V}_p}(\eta,p)}{N e^2\sqrt{\sigma_L+\sigma_R}}
& \displaystyle{\mathop{\sim}_{\eta\to\infty}} &
2^{3/2} \pi\eta \left( p - \frac{A}{1+A} \right)^2 \nonumber \\
& & + c \left[ (1-p)^{3/2} + p^{3/2} \right] . \label{EVinfty}
\end{eqnarray}  
According to (\ref{dimpress}), the dimensionless pressure behaves as
\begin{equation} \label{PressureVx}
\widetilde{P}_0(\eta,p) =
- (1+A)^2 \left( p - \frac{A}{1+A} \right)^2   
\end{equation}
and the function $K_0(\eta,p)$ (\ref{Kgs}) is constant:
\begin{equation}
K_0(\eta,p) = \frac{1+A}{A} .
\end{equation}
The effective fields
\begin{eqnarray}
\kappa_L(\eta,p) & = & 1 - A + p(1+A) , \label{effectfield1} \\
\kappa_R(\eta,p) & = & 2 - p \left( \frac{1+A}{A} \right)  
\label{effectfield2} \end{eqnarray}
depend only on the  {order} parameter $p$.

For fixed values of the parameters $(\eta,A)$, the ground-state value of $p$
is determined by the condition of the energy minimum
(\ref{energymin}) as follows
\begin{equation}
2^{5/2} \pi \eta \left( p_{\rm gs} - \frac{A}{1+A} \right)
+ \frac{3}{2} c \left( \sqrt{p_{\rm gs}} - \sqrt{1-p_{\rm gs}} \right) = 0 . 
\end{equation}
Consequently, at large $\eta$,
\begin{equation} \label{xgsasym}
p_{\rm gs} \mathop{\sim}_{\eta\to\infty} p_{\rm neutr} -
\frac{3 (-c)}{2^{7/2}\pi} \frac{1-\sqrt{A}}{\sqrt{1+A}} \frac{1}{\eta} .  
\end{equation}
Since $c$ is negative, $p_{\rm gs}$ tends to its asymptotic value
$p_{\rm neutr} = A/(1+A)$ from below; this means that the number of
counterions on the right plate $N_R<\sigma_R S$ and the number of
counterions on the left plate $N_L>\sigma_L S$.
The ground-state energy of phase ${\rm V}_p$ has the large-$\eta$
asymptotic
\begin{eqnarray}
\frac{E_{\rm gs}(\eta)}{N e^2\sqrt{\sigma_L+\sigma_R}}
& \displaystyle{\mathop{\sim}_{\eta\to\infty}} &
c \frac{1 + A^{3/2}}{(1+A)^{3/2}} \nonumber \\
& & - \frac{9 c^2}{2^{11/2}\pi} \frac{(1-\sqrt{A})^2}{1+A} \frac{1}{\eta} .  
\end{eqnarray}  
Inserting into (\ref{PressureVx}) $p=p_{\rm gs}$ from (\ref{xgsasym}),
the ground-state pressure exhibits the following asymptotic behavior
\begin{equation} \label{aspress}
\widetilde{P}_{\rm gs}(\eta) \mathop{\sim}_{\eta\to\infty}
- \frac{9 c^2}{2^7 \pi^2} (1-\sqrt{A})^2 (1+A) \frac{1}{\eta^2} .  
\end{equation}
This formula is non-universal because it contains the Madelung constant
$c$ of the hexagonal Wigner structure and the asymmetry parameter $A$.
As $\widetilde{P}_{\rm gs}$ goes at asymptotically large distances to 0
from below, the asymmetric plates attract one another. 

In the symmetric case $A=1$, the leading long-range $1/\eta^2$ term in
(\ref{aspress}) vanishes and the standard short-range exponential
attraction \cite{Samaj12}
\begin{equation} 
\widetilde{P}_{\rm gs}(\eta) \mathop{\sim}_{\eta\to\infty}
- 3 \exp\left( -\frac{4\pi\eta}{\sqrt{2}3^{1/4}} \right)
\end{equation}
takes place.
This results does not contradict the previous formula
since exponentially decaying contributions were neglected
in the derivation of (\ref{aspress}).

\renewcommand{\theequation}{4.\arabic{equation}}
\setcounter{equation}{0}

\section{Nonzero temperatures} \label{Sec4}
The system is considered to be in thermal equilibrium at
the inverse temperature $\beta=1/(k_{\rm B}T)$.
Besides the dimensionless distance $\eta$ (\ref{eta}) introduced
in the ground state, there are two other length scales relevant for
nonzero temperatures.
The Bjerrum length $\ell_{\rm B}$ is the distance at which
two unit charges interact with the thermal energy $k_{\rm B}T$,
$\ell_{\rm B}=\beta e^2$.
Respecting our convention, the Gouy-Chapman length $\mu$ is the distance
from the left plate at which the potential energy induced by the surface
charge density $e\sigma_L$ equals to the thermal energy $k_{\rm B}T$,
\begin{equation}
\mu = \frac{1}{2\pi\ell_{\rm B}\sigma_L} .
\end{equation}
The perpendicular $z$-coordinate will be expressed in units of $\mu$:
\begin{equation}
\widetilde{z} \equiv \frac{z}{\mu} .
\end{equation}  
The dimensionless coupling parameter $\Xi$, reflecting the strength
of electrostatic correlations, is defined as the ratio of the two
length scales,
\begin{equation}
\Xi \equiv \frac{\ell_{\rm B}}{\mu} = 2\pi\ell_{\rm B}^2 \sigma_L .  
\end{equation}
The dimensionless distance $\eta$ (\ref{eta}) is expressible in terms
of $\widetilde{d}=d/\mu$ as follows
\begin{equation}
\eta = \sqrt{1+A} \frac{\widetilde{d}}{2\sqrt{\pi\Xi}} .
\end{equation}  

Similarly as in the ground state, the pressure can be obtained
via either the contact theorem or the thermodynamic
route.

\subsection{The contact pressure} \label{Sec41}
At nonzero yet not too large temperatures, the smear of the (left or right)
ionic layer due to thermal noise is much smaller than the inter-ion spacing
within the given layer.
One can thus adopt the single-particle ground-state picture with the
precisely same effective fields acting on counterions close to the left
and right plates.
There is an effective electric field $\kappa_L E_L$ [with $E_L$ given by
(\ref{EL})] acting on counterions attached to the left plate,
the corresponding potential reads as $V_L(z) = - \kappa_L E_L z$.
Thermal equilibrium at nonzero temperature is turned on via
the position-dependent counterion density in space $n_L(z)$ which
is proportional to the one-body Boltzmann factor
$\exp\left[ -\beta (-e)V_L(z) \right]$:
\begin{equation}
n_L(z) = C_L \exp\left( -\kappa_L \widetilde{z}\right) .
\end{equation}  
 {The 2D surface density of counterions $n_L$ (which has
dimension $1/({\rm length})^2$) is given as the integral along
the perpendicular $z$-axis of the density of counterions in 3D space
$n_L(z)$ (which has dimension $1/({\rm length})^3$):  
\begin{equation}
n_L = \int_0^d {\rm d}z\, n_L(z) . 
\end{equation}
This normalization condition determines the prefactor $C_L$ as follows
\begin{equation}
C_L = \frac{\kappa_L n_L}{\mu} \frac{1}{1-{\rm e}^{-\kappa_L\widetilde{d}}} .
\end{equation}}
Analogously, the spatial density of counterions attached to
the right plate reads as
\begin{equation}
n_R(z) = C_R \exp\left[ -\kappa_R A (\widetilde{d}-\widetilde{z}) \right]
\end{equation}  
where the prefactor $C_R$ is determined by the normalization condition
$n_R = \int_0^d {\rm d}z\, n_R(z)$ as follows
\begin{equation}
C_R = \frac{\kappa_R n_R A}{\mu} \frac{1}{1-{\rm e}^{-\kappa_R A\widetilde{d}}} .
\end{equation}
The total density of counterions in space $n(z)$ is the sum of the left
and right counterion densities:
\begin{equation}
n(z) = n_L(z) + n_R(z) .  
\end{equation}

The  {order} parameter $p$, introduced in the analytic
treatment of asymmetric plates, cannot be deduced from numerical simulations
at nonzero temperatures because it is not clear from the actual position
of  {a counterion} which is its plate of origin at
zero temperature. 
On the other hand, the knowledge of the density profile in simulations
motivates us to employ the half-space occupation quantity $\tau$
as the ratio of the number of counterions to the right of the midplane
between the plates, $N^>$, to the total counterion number $N$,
\begin{equation} \label{tau}
\tau = \frac{N^>}{N} .
\end{equation}
 {This quantity has already been introduced in \cite{Bazant04}
under the name ``the total diffuse charge near the cathode''.}
Since $N^>=N_L^> + N_R^>$ where
\begin{equation}
N_L^> = S \int_{d/2}^d {\rm d}z\, n_L(z) = S C_L \mu
\int_{\tilde{d}/2}^{\tilde{d}} {\rm d}z\, {\rm e}^{-\kappa_L\tilde{z}}
\end{equation}
and
\begin{equation}
N_R^> = S \int_{d/2}^d {\rm d}z\, n_R(z) = S C_R \mu
\int_{\tilde{d}/2}^{\tilde{d}} {\rm d}z\, {\rm e}^{-\kappa_RA(\tilde{d}-\tilde{z})} ,  
\end{equation}
the theoretically predicted value of $\tau$ takes the explicit form
\begin{equation}
\tau(\eta,p) = (1-p) \frac{1}{{\rm e}^{\kappa_L\tilde{d}/2}+1} +
p \frac{1}{{\rm e}^{-\kappa_R A\tilde{d}/2}+1} .  
\end{equation}
The limiting $\eta\to 0$ and $\eta\to\infty$ values of $\tau$ coincide
with those of $p$:
\begin{equation} \label{bctau}
\lim_{\eta\to 0} \tau = \frac{1}{2} , \qquad
\lim_{\eta\to\infty} \tau = \frac{A}{1+A} .
\end{equation}  

Having the density profile of counterions, the pressure $P_c$ can be obtained
by applying the contact value theorem for planar wall surfaces
\cite{Henderson78,Henderson79,Carnie81,Wennerstrom82}.
With respect to the left plate, the pressure is given by
\begin{eqnarray}
\beta P_c & = & n(0) - 2\pi\ell_{\rm B} \sigma_L^2 \nonumber \\
& = & C_L + C_R \exp\left( -\kappa_R A \widetilde{d}\right)  
- 2\pi\ell_{\rm B} \sigma_L^2 .
\end{eqnarray}
Introducing the dimensionless pressure 
\begin{equation} \label{widetildeP}
\widetilde{P}_c\equiv \frac{\beta P_c}{2\pi\ell_{\rm B}\sigma_L^2} ,
\end{equation}
one gets that
\begin{equation} \label{P1}
\widetilde{P}_c = \frac{\kappa_L n_L}{\sigma_L}
\frac{1}{1-{\rm e}^{-\kappa_L\widetilde{d}}} + \frac{\kappa_R n_R}{\sigma_R}
A^2 \frac{1}{{\rm e}^{\kappa_R A\widetilde{d}}-1} -1 .
\end{equation}
With respect to the right plate, the pressure is given by
\begin{eqnarray}
\beta P_c & = & n(d) - 2\pi\ell_{\rm B} \sigma_R^2 \nonumber \\
& = & C_L \exp\left( -\kappa_L \widetilde{d}\right) + C_R 
- 2\pi\ell_{\rm B} \sigma_R^2 ,
\end{eqnarray}  
so that
\begin{equation} \label{P2}
\widetilde{P}_c = \frac{\kappa_L n_L}{\sigma_L}
\frac{1}{{\rm e}^{\kappa_L\widetilde{d}}-1} + \frac{\kappa_R n_R}{\sigma_R} A^2
\frac{1}{1-{\rm e}^{-\kappa_R A\widetilde{d}}} - A^2 .
\end{equation}

The requirement of the equivalence of the two pressure representations
(\ref{P1}) and (\ref{P2}),
\begin{equation}
\kappa_L(1-p) - \kappa_R A p = 1-A ,  
\end{equation}
coincides with the ground-state constraint for the left and right effective
fields (\ref{equalprime}).

In analogy with the ground-state Eqs. (\ref{kappaL}) and (\ref{kappaR}),
this constraint is fulfilled by the ansatz
\begin{eqnarray}
\kappa_L(\eta,p) & = & 1 - A + A p K(\eta,p) , \label{kappaLL} \\
\kappa_R(\eta,x) & = & 1 - \frac{1}{A} + (1-p) K(\eta,p) \label{kappaRR}
\end{eqnarray}
with $K(\eta,x)$ being an arbitrary function.
The basic idea of the theory in \cite{Palaia22} dealing with the symmetric
case was that extending the ground-state effective fields to nonzero
temperatures is a plausible approximation.
The same idea is adopted to our asymmetric case by setting
\begin{equation}
K(\eta,p) = K_0(\eta,p) ,
\end{equation}
with $K_0(\eta,p)$ defined by (\ref{Kgs}),
in the above relations (\ref{kappaLL}) and (\ref{kappaRR}).
In this way one obtains the ground-state representations
(\ref{kappaL}) and (\ref{kappaR}).

Using the formulas (\ref{P1}) and (\ref{P2}) for the effective fields,
the pressure can be expressed in a symmetrized form:
\begin{eqnarray}
\widetilde{P}_c & = & \frac{1}{2} \left[ \frac{\kappa_L n_L}{\sigma_L}
\coth\left(\frac{\kappa_L\widetilde{d}}{2}\right)
+ \frac{\kappa_R n_R}{\sigma_R} A^2
\coth\left(\frac{\kappa_R A\widetilde{d}}{2}\right) \right] \nonumber \\
& & - \frac{1+A^2}{2} . \label{pressure}
\end{eqnarray}  
Applying here the expansion formula
\begin{equation}
\coth t = \frac{1}{t} + \frac{t}{3} + {\cal O}(t^3) ,
\end{equation}
the limiting $\eta\to 0$ values of $\kappa_L$ (\ref{limits1})
and $\kappa_R$ (\ref{limits2}) and the constraint (\ref{equal}), 
the small-distance expansion of the pressure is expressible explicitly
up to the first order in $\widetilde{d}$:
\begin{equation}
\widetilde{P}_c = \frac{1+A}{\widetilde{d}} - \frac{1+A^2}{2} +
(1-A)^2 (1+A) \frac{\widetilde{d}}{12} + {\cal O}(\widetilde{d}^2) . 
\end{equation}  

\subsection{The pressure obtained via the thermodynamic route} \label{Sec42}
In the canonical ensemble, having $N_L$ counterions
attached to the left plate and $N_R$ counterions attached
to the right plane, the free energy $F(N_L,N_R)$ is defined as
\begin{equation} \label{freeenergy}
- \beta F(N_L,N_R) = \ln Z(N_L,N_R) ,
\end{equation}
where $Z(N_L,N_R)$ is the partition function
\begin{eqnarray} 
Z_N & = & \frac{1}{(N_L+N_R)!} \int_{\Lambda} \prod_{j=1}^{N_L} 
\frac{{\rm d}{\bf r}_j}{\lambda^3}
\prod_{k=1}^{N_R} \frac{{\rm d}{\bf r}_k}{\lambda^3} \nonumber \\ & & \times
\exp\left[ -\beta E(\eta,x;\{ {\bf r}_j\},\{ {\bf r}_k\}) \right] ,
\label{generalpartition} 
\end{eqnarray}
where $\lambda$ stands for the thermal de Broglie wavelength and
$\Lambda = \{ {\bf r},0<z<d \}$ denotes the slab between the plates.

The total energy of counterions $E(\eta,x;\{ {\bf r}_j\},\{ {\bf r}_k\})$
can be expanded around the ground state energy in small deviations from their
ground state positions, as is done for symmetrically charged plates in
Ref. \cite{Samaj11b}, 
\begin{eqnarray} 
\beta E(\{ {\bf r}_j\},\{ {\bf r}_k\}) & = & \beta E_{\rm gs} 
+ \kappa_L \sum_{j=1}^{N_L} \tilde{z}_j
+ \frac{\ell_{\rm B}}{2 a_L^3} \sum_{j=1}^{N_L} \left( x_j^2+y_j^2 \right)
\nonumber \\ & &
+ \kappa_R A \sum_{k=1}^{N_R}(\tilde{d}-\tilde{z}_k) 
\nonumber \\ & &
+ \frac{\ell_{\rm B}}{2 a_R^3} \sum_{k=1}^{N_R} \left( x_k^2+y_k^2 \right)
+ \cdots . \label{generalenergy}
\end{eqnarray}
Here, $a_L\propto 1/\sqrt{n_L}$ and $a_R\propto 1/\sqrt{n_R}$ are the lattice
spacings of  {counterion} structures created on the left and
right plates, respectively; the prefactors, which depend only on the particular
lattice structures, are irrelevant for our purposes.
Higher-order terms in (\ref{generalenergy}) scale like
$1/\Xi^{(a-2)/4}$ $(a=3,4\ldots)$ and therefore vanish in the limit
$\Xi\to\infty$.

The integration of the Boltzmann factor with the energy
(\ref{generalenergy}) can be performed straightforwardly in
(\ref{generalpartition}) for the perpendicular $z$-components:
\begin{equation}
\int_0^d \frac{{\rm d}z_j}{\lambda} {\rm e}^{-\kappa_L\tilde{z}_j} =
\frac{\mu}{\lambda} \frac{1-{\rm e}^{-\kappa_L\tilde{d}}}{\kappa_L} ,
\quad j=1,\ldots, N_L 
\end{equation}
and
\begin{equation}
\int_0^d \frac{{\rm d}z_k}{\lambda} {\rm e}^{-\kappa_R A(\tilde{d}-\tilde{z}_k)} =
\frac{\mu}{\lambda} \frac{1-{\rm e}^{-\kappa_R A\tilde{d}}}{\kappa_R A} ,
\quad k=1,\ldots, N_R .
\end{equation}
The parallel components $x,y$ are trickier.
Let us assume first that due to strong electrostatic repulsions in
the parallel $(x,y)$ plane a given counterion
(say the one sitting on the left plate) is constrained to the space
$S/N_L=1/n_L$ reserved for one counterion,
namely to a disk of radius $R_L^2=1/(\pi n_L)$.
In radial coordinates, the integration over coordinates $x,y$ reads as
\begin{equation} \label{integr}
\int_0^{R_L} 2\pi r {\rm d}r \exp\left( -\frac{\ell_{\rm B}}{2 a_L^3} r^2 \right)
= \frac{2\pi a_L^3}{\ell_{\rm B}} \int_0^{\frac{\ell_{\rm B}}{2\pi n_L a_L^3}}
{\rm d}t\, {\rm e}^{-t} .
\end{equation}
Since the upper limit of integration 
\begin{equation}
\frac{\ell_{\rm B}}{2\pi n_L a_L^3} \propto \frac{\ell_{\rm B}}{a_L}
\propto \sqrt{\Xi} \sqrt{\frac{n_L}{\sigma_L}} ,  
\end{equation}
the integral over $t$ in (\ref{integr})
equals to 1 in the large-$\Xi$ limit. 
Consequently, the integration over coordinates $x,y$ implies for every
 {counterion} the factor $\propto n_L^{-3/2}$ where
the explicit form of the constant prefactor is irrelevant.
Similarly, the corresponding factor for each  {counterion}
sitting on the right plate can be shown to be $\propto n_R^{-3/2}$. 

To summarize the above paragraph, the free energy (\ref{freeenergy})
can be expressed as
\begin{eqnarray}
-\beta F(N_L,N_R) & = & {\rm const} - \beta E_{\rm gs}
\nonumber \\ & &
+ N_L \ln\left( \frac{1-{\rm e}^{-\kappa_L\tilde{d}}}{\kappa_L} \right) 
- \frac{3}{2} N_L \ln n_L
\nonumber \\ & &
+ N_R \ln\left( \frac{1-{\rm e}^{-\kappa_R A\tilde{d}}}{A\kappa_R} \right) 
- \frac{3}{2} N_R \ln n_R . \nonumber \\ & &
\end{eqnarray}
Thus, the free energy per  {counterion}
$f = F/(N_L+N_R)$ is given by
\begin{eqnarray}
\beta f(\eta,p) & = & {\rm const} + \frac{\beta E_{\rm gs}(\eta,p)}{N}
- (1-p) \ln\left( \frac{1-{\rm e}^{-\kappa_L\tilde{d}}}{\kappa_L} \right) 
\nonumber \\ & & + \frac{3}{2} (1-p) \ln (1-p)
- p \ln\left( \frac{1-{\rm e}^{-\kappa_R A\tilde{d}}}{A\kappa_R} \right) 
\nonumber \\ & & + \frac{3}{2} p \ln p . \label{freeenergytotal}
\end{eqnarray}
The thermodynamic pressure is defined as (minus) the total derivative of
the free energy with respect to the distance:
\begin{eqnarray} 
P_{th}(\eta,p) & = & - \frac{\rm d}{{\rm d} d} \frac{F(\eta,p)}{S}
\nonumber \\ & = &
- \frac{1}{\sqrt{2}} (\sigma_L+\sigma_R)^{3/2}
\frac{{\rm d} f(\eta,p)}{{\rm d}\eta} . \label{thermopress}
\end{eqnarray}
The value of the  {order} parameter $p$ is determined by
the variational condition of the free energy minimum:
\begin{equation} \label{freeenergymin}
\frac{\partial f(\eta,x)}{\partial x} = 0 , \qquad
\frac{\partial^2 f(\eta,x)}{\partial x^2} < 0 .  
\end{equation}
The dimensionless thermodynamic pressure
$\widetilde{P}_{th}\equiv \beta P_{th}/(2\pi\ell_{\rm B}\sigma_L^2)$
is thus expressible as  
\begin{equation} \label{dimthermopress}
\widetilde{P}_{th}(\eta,p) = - \frac{1}{2^{3/2}\pi} (1+A)^2
\frac{\partial}{\partial \eta}
\frac{\beta f(\eta,p)}{\ell_{\rm B}\sqrt{\sigma_L+\sigma_R}} ,
\end{equation}
where the replacement of the total derivative by the partial one is possible
due to the stationarity condition (\ref{freeenergymin}). 

As soon as the temperature is nonzero, i.e., for any finite value of
the coupling constant $\Xi$, the effective force (pressure) between
two symmetrically charged walls at asymptotic (in fact, extremely large)
distances between the walls $d\to\infty$ is expected to be {\em repulsive},
of the PB power-law type
\cite{Netz01,Shklovskii99,Chen06,Santangelo06,dosSantos09}
\begin{equation} \label{PBpressure}
\beta P \mathop{\sim}_{d\to\infty} \frac{\pi}{2\ell_{\rm B}} \frac{1}{d^2} ,
\qquad \widetilde{P} \mathop{\sim}_{\tilde{d}\to\infty}
\frac{\pi^2}{\tilde{d}^2} .  
\end{equation}
The validity of this asymptotic relation for asymmetric plates (but with the same sign of the charge) was shown in Ref. \cite{Kanduc08}.
Note the universal independence of the repulsive pressure (\ref{PBpressure})
on the surface charge densities of the plates, the only condition is that
the plates bear surface charges of the same sign.
For asymmetric plates, the scaling (\ref{PBpressure}) is fulfilled at larger
distances compared to symmetric plates.

As was shown in the previous section, at zero temperature phase $V_p$ is
relevant at asymptotically large distances, leading to an {\em attractive}
pressure with the asymptotic non-universal behavior (\ref{aspress})
which has the same $1/d^2$-dependence on the distance as the PB pressure
(\ref{PBpressure}).
At  {low} enough temperatures, the attractive regime still
exists and spreads over a large interval of distances, except for
asymptotically large distances where the PB repulsion takes place.
One can intuitively expect the impact of phase $V_x$ on the large-distance
behavior of the attractive pressure within the given interval of distances.
To be more particular, let us consider the ground state energy (\ref{EVinfty})
with the effective fields (\ref{effectfield1}) and (\ref{effectfield2})
in the expression for the free energy (\ref{freeenergytotal}). 
The exponentially decaying terms of type $\exp(-\kappa_L \tilde{d})$
and $\exp(-\kappa_R A \tilde{d})$ are negligible comparing with
their inverse-power-law counterparts, thus the free energy is given by
\begin{eqnarray}
\frac{\beta f(\eta,p)}{\ell_{\rm B}\sqrt{\sigma_L+\sigma_R}} & = &
{\rm const} + 2^{3/2}\pi\eta \left( p-\frac{A}{1+A} \right)^2
\nonumber \\  & & + c \left[ (1-p)^{3/2} + p^{3/2} \right]
+ \sqrt{\frac{2\pi}{(1+A)\Xi}} \nonumber \\  & &
\times \Big\{ (1-p) \ln[1-A+p(1+A)]  \nonumber \\  & &
+ \frac{3}{2} (1-p) \ln(1-p) \nonumber \\  & & 
+ p \ln[2A-p(1+A)] + \frac{3}{2} p\ln p \Big\} . \nonumber \\  & &  
\end{eqnarray}  
The variational condition (\ref{freeenergymin}) implies that for large $\eta$
\begin{equation} \label{a1}
p\sim \frac{A}{1+A} + \frac{a}{\eta} , \quad
a = \frac{3 c}{2^{7/2}\pi} \frac{1-\sqrt{A}}{\sqrt{1+A}}
- \frac{5}{8\sqrt{\pi\Xi}} \frac{\ln A}{\sqrt{1+A}} .
\end{equation}
In the zero-temperature limit $\Xi\to\infty$, this formula reduces to
the previous ground-state one (\ref{aspress}) as it should be.
The thermodynamic pressure (\ref{dimthermopress}) behaves for large
$\eta$ as follows
\begin{equation} \label{a2}
\widetilde{P}_{th} = - (1+A)^2\left( p-\frac{A}{1+A} \right)^2
\sim - (1+A)^2 \frac{a^2}{\eta^2} .
\end{equation}
This scaling dependence on the distance between the plates, which holds
exclusively in the large-distance region of the attractive pressure,
has the functional form of the repulsive PB pressure (\ref{PBpressure}),
with a non-universal prefactor.

\renewcommand{\theequation}{5.\arabic{equation}}
\setcounter{equation}{0}

\section{Monte Carlo simulations} \label{Sec5}
Metropolis MC simulations were carried out in a quasi-2D slab
geometry, where $x$ and $y$ directions are periodic.
The last $z$-direction is bound by two charged, planar, and hard surfaces,
with uniform surface charge densities $\sigma_L$ and $\sigma_R$.
We used $N=384$ mobile point charges, which neutralize the surface charges,
and varied both the electrostatic coupling parameter $\Xi$ and separation $d$
between the two charged plates.
The point charges were confined to the slab between the two surfaces.
Electrostatic interactions were handled with standard Ewald summation
techniques, where we introduced an extra vacuum slab between the periodic
images in the $z$-direction, with corrections for the quasi-2D dimensionality
and the extra vacuum slab \cite{Yeh99, Mazars01}.
The correction term in our case (keeping only terms dependent on
the mobile charges positions) equals:
\begin{eqnarray}
\beta U_{q2D} & = & 2 \pi l_B  \Bigg[ \frac{1}{(d+v)S} \left(
\sum_i q_i z_i \right)^2 \nonumber \\ & &
+ \frac{(\sigma_L+\sigma_R)}{(d+v)} \sum_i q_i z_i^2
+ (\sigma_R-\sigma_L) \sum_i q_i z_i \Bigg], \nonumber \\ & & 
\end{eqnarray}
where $v$ is the length of the vacuum slab, $S$ the area of either of
the surfaces, $z_i$ the perpendicular position of the charges (where
the midplane is defined as $z=0$), and $q_i$ the valency of charge $i$.  
The vacuum slab was usually set to be $v=200\mu$ wide.
Tests with larger slabs were performed but without any detectable difference.
We also varied the precision of the Ewald summation, including more terms
in the Fourier space summation and faster damping of the real part,
but again, without any detectable differences compared to the reported data. 
New trial configurations were generated by randomly displacing them
a certain distance, with an acceptance ratio close to around 30-50\%
using the Metropolis MC algorithm.
For a tenth of these displacements, we also tried to mirror a point charge
around the midplane to the other surface.
All data point were pre-equilibrated for $10^4$ MC cycles,
where one cycles corresponds to $N$ trial displacements.
Pressures and ion density profiles were then collected over $10^5$ MC cycles. 
Pressures were either calculated by estimating the contact value of
the ion densities at the respective surface, minus either $2 \pi \sigma_L^2$
or $2 \pi \sigma_R^2$ accounting for the electrostatic interaction between
the smeared-out charge of the surfaces with the rest of the system,
or over the midplane.
The latter involves both estimating the midplane concentration and all
the electrostatic forces acting across the midplane \cite{Guldbrand84}.
While both estimates of the pressures are the same, the latter is usually
more precise.
Standard errors in pressure were estimated by applying block-averages,
using ten blocks.

\renewcommand{\theequation}{6.\arabic{equation}}
\setcounter{equation}{0}

\section{Comparison of the theory with Monte Carlo data} \label{Sec6}
All MC simulations were done with the asymmetry parameter $A=\frac{1}{2}$.
The thermodynamic quantities of interest are the (dimensionless) pressure
$\widetilde{P}$ (\ref{widetildeP}) and the half-space occupation parameter
$\tau$ defined by (\ref{tau}). 

For very large values of the coupling constant $\Xi>300$, the results of
the theory agree remarkably with MC data; to spare space we do not
present them.

\begin{figure}[tbp]
\begin{center}
\includegraphics[clip,width=0.48\textwidth]{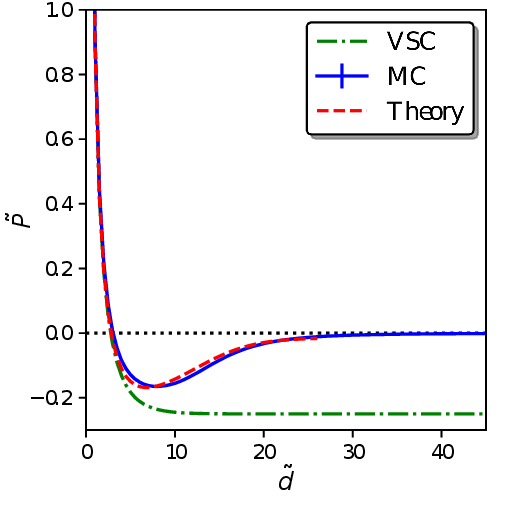}
\caption{The pressure $\widetilde{P}$ versus the distance $\tilde{d}$ for
the asymmetry parameter $A=\frac{1}{2}$ and the coupling constant $\Xi=100$.
Solid line corresponds to MC data, dashed line to the present theory
with the pressure obtained via the thermodynamic route and dash-and-dot
line to the leading term of the VSC theory \cite{Kanduc08}, see (\ref{PSC}).}
\label{fig2}
\end{center}
\end{figure}

The results for the pressure $\widetilde{P}$ as the function of
the distance $\tilde{d}$ for the intermediate value of the coupling
constant $\Xi=100$ are presented in Fig. \ref{fig2}.
The MC data are represented by solid line and the results of
the present theory by dashed line; the thermodynamic route to obtain
the pressure provides more reliable results than the contact value theorem,
so theoretical results are taken for that pressure.
The results of the present theory are limited to the range of distances
$0<\tilde{d}\lesssim 25$, where the ground state corresponds to the phases I
and ${\rm I}_x$; for larger distances where the phase ${\rm V}_x$ prevails,
see the large-distance analysis of the attracive pressure at the end of
Sec.~\ref{Sec4}.
It is seen that the agreement of the theory and MC simulations is excellent
also for this not too-large value of $\Xi$.
The leading term of the virial SC (VSC) theory for the asymmetrically
charged plates \cite{Kanduc08}, given by
\begin{equation} \label{PSC}
\widetilde{P}_{\rm VSC} = - \frac{1}{2} \left( 1 + A^2 \right)
+ \frac{1}{2} \left( 1 - A^2 \right)
\coth\left( \frac{1-A}{2} \tilde{d} \right) ,   
\end{equation}
is represented by the dash-and-dot curve.
The VSC theory gives reasonable values of the pressure only for
small values of $\tilde{d}$ corresponding to the repulsive regime of
the pressure.  

\begin{figure}[tbp]
\begin{center}
\includegraphics[clip,width=0.48\textwidth]{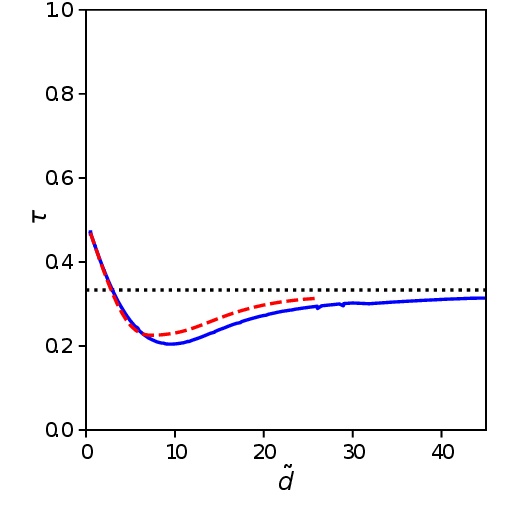}
\caption{The half-space occupation parameter $\tau$ versus the distance
$\tilde{d}$ for the asymmetry parameter $A=\frac{1}{2}$ and
the coupling constant $\Xi=100$.
Solid line corresponds to MC data and dashed line to the present theory.
The horizontal dotted line shows the large-distance asymptotics as given by
Eq. (\ref{bctau}), yielding here $\tau=\frac{1}{3}$.}
\label{fig3}
\end{center}
\end{figure}

The results for the half-space occupation parameter $\tau$ as the function of
the distance $\tilde{d}$ for the coupling constant $\Xi=100$
are presented in Fig. \ref{fig3}.
As before, the MC data are represented by solid line and the results of
the present theory by dashed line.
The agreement of the theory and MC simulations is very good as well.
The limiting $\eta\to 0$ and $\eta\to\infty$ values of $\tau$ satisfy
the requirements (\ref{bctau}).
It is interesting that the plot of $\tau(\tilde{d})$ is not monotonous.
For very small distances between the walls the numbers of
counterions attached to the left and right walls are equal,
then increasing the distance counterions migrate from
the right to the left walls.
The half-space occupation parameter is minimal when counterions
maximally attach to the left wall. For larger $d$, beyond this minimum, 
counterions move to the right.
For infinite d, counterions fully screen the surface charge of both plates.

\begin{figure}[tbp]
\begin{center}
\includegraphics[clip,width=0.48\textwidth]{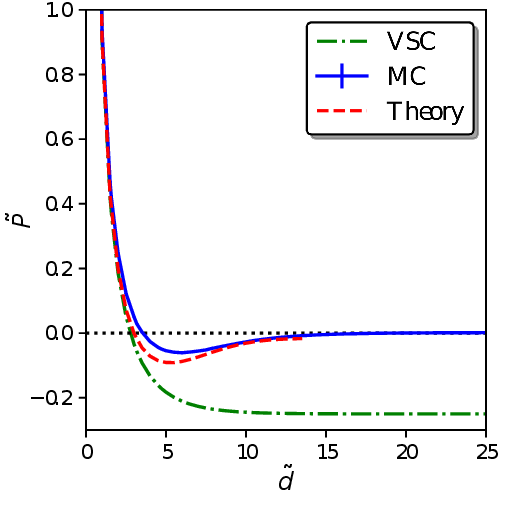}
\caption{The pressure $\widetilde{P}$ versus the distance $\tilde{d}$ for
the asymmetry parameter $A=\frac{1}{2}$ and the coupling constant $\Xi=30$.
The notation is the same as in Fig. \ref{fig2}.}
\label{fig4}
\end{center}
\end{figure}

\begin{figure}[tbp]
\begin{center}
\includegraphics[clip,width=0.48\textwidth]{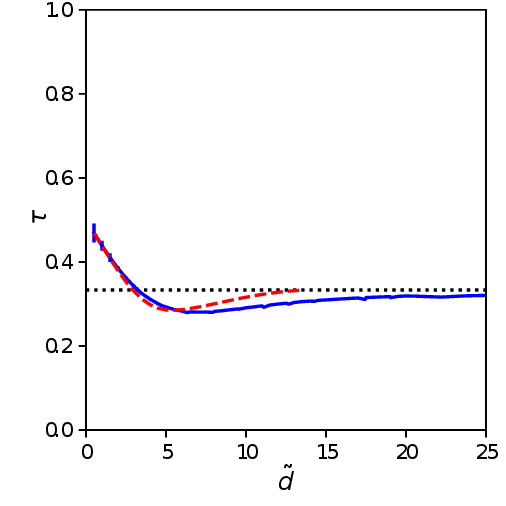}
\caption{The half-space occupation parameter $\tau$ versus the distance
$\tilde{d}$ for the asymmetry parameter $A=\frac{1}{2}$ and
the coupling constant $\Xi=30$.
The notation is the same as in Fig. \ref{fig3}.}
\label{fig5}
\end{center}
\end{figure}

The results for the pressure $\widetilde{P}$ and the half-space occupation
parameter $\tau$ as the functions of the distance $\tilde{d}$ for the coupling
constant $\Xi=30$ are presented in Figs. \ref{fig4} and \ref{fig5},
respectively.
The agreement of the theory and MC simulations is very good also for this
value of $\Xi$.

Let us now discuss the asymptotic large-distance behavior of the pressure
$\widetilde{P}$ observed in MC simulations.
For large values of the coupling constant $\Xi$, there is an apparent
scaling regime for the {\em attractive} $\widetilde{P}<0$ in the region
of large (but not too large) distances as predicted by
the relations (\ref{a1}) and (\ref{a2}).
At the same time, the PB regime (\ref{PBpressure}) of the {\em repulsive}
$\widetilde{P}>0$ takes place at extremely large distances which are
usually not accessible to standard MC simulations due to the lack of accuracy.
The scaling region for the attractive $\widetilde{P}$ becomes less
pronounced when decreasing $\Xi$ and it even disappears for small values
of $\Xi$.
As concerns the scaling PB region for the repulsive $\widetilde{P}$,
it moves down to smaller distances when decreasing $\Xi$ and for small
values of $\Xi$ it is readily accessible by using standard MC simulations.

\begin{figure}[tbp]
\begin{center}
\includegraphics[clip,width=0.48\textwidth]{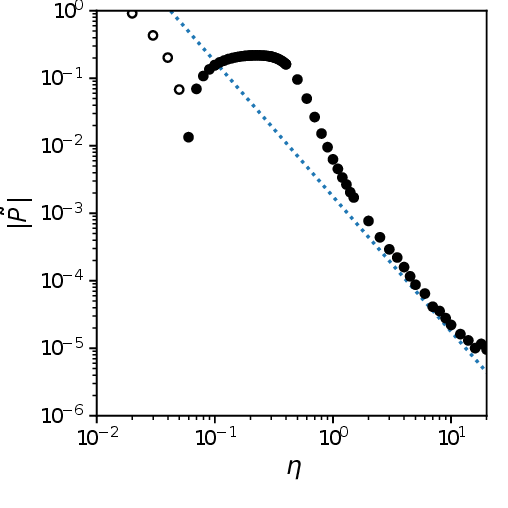}
\caption{The log-log plot of the absolute value of the pressure
$\vert \widetilde{P}\vert$ versus the distance $\eta$ for the coupling
constant $\Xi=300$.
MC data are represented by open/filled circles for repulsive/attractive forces.
The theoretical prediction of the asymptotic behavior (\ref{theory})
is represented by dotted line.}
\label{fig6}
\end{center}
\end{figure}

To be more particular, the log-log plot of the absolute value of the pressure
$\vert \widetilde{P}\vert$ versus the distance $\eta$
for the coupling constant $\Xi=300$ is pictured in Fig. \ref{fig6}.
MC data are represented by open symbols (circles) for repulsive forces
(pressures) and filled ones for attractive forces.
The vacuum gap is here $800\mu$ units. 
Dotted line is the theoretical prediction of the asymptotic behavior
of the attractive pressure
\begin{equation} \label{theory}
\widetilde{P}_{th} \sim - 0.00177046... \frac{1}{\eta^2}
\end{equation}  
obtained from the relations (\ref{a1}) and (\ref{a2}) taken at $A=\frac{1}{2}$
and $\Xi=300$.
The agreement between MC data and this prediction is remarkable.  
 {It is seen that for the coupling constant $\Xi=300$
the scaling regime for the attractive pressure (\ref{theory}) 
starts at $\eta\approx 3$ ;  where it stops is hard to judge, as we lose numerical precision (for the MC simulations) around $\eta=10$ for this $\Xi$-value.}   
We add that at the ground state characterized by $\Xi\to\infty$, the
asymptotic behavior of the pressure for $A=\frac{1}{2}$ takes the form
with a quite distinct (almost doubled) prefactor:
\begin{equation}
\widetilde{P}_{th} \sim - 0.00352357... \frac{1}{\eta^2} .
\end{equation}  

\renewcommand{\theequation}{7.\arabic{equation}}
\setcounter{equation}{0}

\section{Conclusion} \label{Sec7}
In the context of the effective interaction between symmetrically charged
parallel plates mediated by counterions, it was shown in Ref. \cite{Palaia22}
that the effective fields created at zero temperature by the plate surface
charges and the lattice structures of counterions on the plates  {are also relevant at nonzero temperatures, and rule the density profiles (hence also rule the pressure).}
The present work extends this effective field method to asymmetrically
charged plates.
The technical complication in the asymmetric problem comes from the fact
that each plate as a whole (i.e., the surface charge density plus the cloud
of counterions attached to that plate) is not neutral.
This causes stronger long-ranged interaction effects between the plates
and requires the introduction of an additional (occupation) order parameter $p$
(\ref{x}) into the theory.
This parameter is defined unambiguously in the ground state (counterions stuck
on the plate surfaces) and its value is determined variationally
to ensure the minimum of the energy (\ref{energymin}).
At nonzero temperatures, the order parameter $p$ represents an auxiliary
variational quantity which ensures the minimum of the free energy
(\ref{freeenergymin}).
Since at $T>0$ the  {order} parameter $p$ cannot be measured
in MC simulations, we have introduced the half-space occupation parameter
$\tau$ (\ref{tau}) whose values are available in simulations. 
The theoretical results for the dimensionless pressure $\widetilde{P}$
and the half-space occupation parameter $\tau$ agree very well with MC data,
for the intermediate coupling constant $\Xi=100$ (see Figs. \ref{fig2} and
\ref{fig3}) as well as $\Xi=30$ (see Figs. \ref{fig4} and \ref{fig5}).
By construction, our treatment improves in accuracy when $\Xi$ is increased.
 {It becomes exact for $\Xi\to\infty$}.

An interesting result following from the present work deals with the
asymptotic scaling behavior of the attractive pressure.
It is known that at nonzero temperatures the pressure is repulsive
at asymptotically large distances between the plates and takes
the universal (i.e., independent of the like surface charge densities on
the plates) Poisson-Boltzmann (PB) form (\ref{PBpressure}).
On the contrary, the asymptotic pressure is attractive at zero temperature,
see the non-universal formula (\ref{aspress}) which contains the Madelung
constant of the hexagonal structure $c$ and the asymmetry parameter $A$.
The attraction phenomenon exists also for nonzero temperatures where
two regions of large distances exist: {\em extremely} large distances at which
the repulsive PB pressure (\ref{PBpressure}) takes place and large distances
at which the attractive pressure prevails, with the finite-$\Xi$ correction
in the non-universal prefactor given by Eqs. (\ref{a1}) and (\ref{a2}).
We are thus in the situation of an intermediate asymptotics.
The MC data of the pressure versus distance for the coupling constant $\Xi=300$ 
in Fig. \ref{fig6} are in perfect agreement with our theoretical prediction
represented by dotted line.
Of course, going to higher temperatures the full region of the attractive
pressure diminishes and finally disappears for high enough temperatures.

Since the present method relies on the effective fields acting in the ground
state, it is applicable to the region of large coupling constants $\Xi$.
The fact that it provides reasonable results for the pressure $\widetilde{P}$
and half-space occupation parameter $\tau$ for a coupling constant $\Xi$ as small as 30,
see Figs. \ref{fig4} and \ref{fig5}, is rather surprising.
It would be useful to establish a theory covering both $\Xi\to\infty$ and $\Xi\to 0$ limits, to see the mutual
interconnection and the corresponding ranges of the attractive and repulsive
scaling regions of the pressure.
A possible candidate is the SC method based on the idea of correlation holes
substituting the Wigner crystal of counterions \cite{Palaia18}.

\begin{acknowledgments}
We acknowledge useful discussions with Ivan Palaia.
L. \v{S}. is grateful to LPTMS for hospitality. 
The support of L. \v{S}. received from VEGA Grant No. 2/0089/24 and
Project APVV-20-0150 is acknowledged.
M. T. thanks the Swedish Research Council for financial support
(Grant no. 2021-04997).
\end{acknowledgments}

\appendix

\renewcommand{\theequation}{A.\arabic{equation}}
\setcounter{equation}{0}

\begin{widetext}
\section{Phase ${\bf I}_p$} \label{appA}
The energy of phase ${\rm I}_p$ with commensurate values of
$p\in \{ 1/2,1/3,1/4,1/7,1/9,\cdots\}$ is given by \cite{Antlanger18}
\begin{equation} \label{appA1}
\frac{E_{{\rm I}_p}(\eta,p)}{N e^2\sqrt{\sigma_L+\sigma_R}} =
2^{3/2} \pi \eta \left( p - \frac{A}{1+A} \right)^2 + c
+ \frac{p}{\sqrt{2}} \left[
- {\cal K}(\eta) +\sqrt{p} {\cal K}(\sqrt{p}\eta) \right] ,
\end{equation}
where
\begin{equation} \label{appA2}
{\cal K}(\eta) = \frac{1}{\sqrt{\pi}} \int_0^{\infty}
\frac{{\rm d}t}{\sqrt{t}} \left( 1 - {\rm e}^{-3 t} \right)
\left\{ \left[ \theta_3({\rm e}^{-\sqrt{3}t}) \theta_3({\rm e}^{-t/\sqrt{3}})
 -1-\frac{\pi}{t} \right] 
+ \left[ \theta_2({\rm e}^{-\sqrt{3}t}) \theta_2({\rm e}^{-t/\sqrt{3}})
-\frac{\pi}{t} \right] \right\} .
\end{equation}
Introducing the generalized Misra functions
\begin{equation} \label{Misra}
z_{\nu}(x,y) = \int_0^{1/pi} \frac{{\rm d}t}{t^{\nu}} {\rm e}^{-xt} {\rm e}^{-y/t},  
\end{equation}
in terms of the functions
\begin{eqnarray}
I_2(x,y) & = & 2 \sum_{j=1}^{\infty} (-1)^j \left[
z_{3/2}\left( x,y+\sqrt{3}j^2\right) +
z_{3/2}\left( x,y+\frac{j^2}{\sqrt{3}} \right) \right]
+ 4 \sum_{j,k=1}^{\infty} (-1)^{j+k}
z_{3/2}\left( x,y+\sqrt{3}j^2+\frac{k^2}{\sqrt{3}}\right) , \nonumber \\
I_3(x,y) & = & 2 \sum_{j=1}^{\infty} \left[ z_{3/2}\left( x,y+\sqrt{3}j^2\right) +
z_{3/2}\left( x,y+\frac{j^2}{\sqrt{3}} \right) \right] + 4 \sum_{j,k=1}^{\infty}
z_{3/2}\left( x,y+\sqrt{3}j^2+\frac{k^2}{\sqrt{3}}\right)
- \pi z_{1/2}(x,y),  \nonumber \\
I_4(x,y) & = & 4 \sum_{j,k=1}^{\infty}
z_{3/2}\left( x,y+\sqrt{3}(j-1/2)^2+\frac{(k-1/2)^2}{\sqrt{3}}\right)
- \pi z_{1/2}(x,y),
\end{eqnarray}  
${\cal K}(\eta)$ can be expressed as
\begin{equation}
{\cal K}(\eta) = \frac{1}{\sqrt{\pi}} \left[
I_2(0,0) - I_2\left( (\pi\eta)^2,0\right) + 2 I_3(0,0)
- I_3\left( (\pi\eta)^2,0\right) - I_3(0,\eta^2) + I_4(0,0)
- I_4(0,\eta^2) \right] . 
\end{equation}  

\renewcommand{\theequation}{B.\arabic{equation}}
\setcounter{equation}{0}

\section{Phase ${\bf V}_p$} \label{appB}
The energy of phase ${\rm V}_p$ with commensurate values of
$p\in \{ 1/2,1/4,1/5,1/8,1/10,\cdots \}$ is given by \cite{Antlanger18}
\begin{equation} \label{EV}
\frac{E_{{\rm V}_p}(\eta,p)}{N e^2\sqrt{\sigma_L+\sigma_R}} =
2^{3/2} \pi\eta \left( p - \frac{A}{1+A} \right)^2
+ c \left[ (1-p)^{3/2} + p^{3/2} \right] + J(\eta,p) , 
\end{equation}  
where
\begin{eqnarray} 
J(\eta,p) & = &  p \sqrt{1-p} \frac{1}{2^{3/2}\sqrt{\pi}}
\int_0^{\infty} \frac{{\rm d}t}{\sqrt{t}} \left[ - {\rm e}^{-\eta^2(1-p)t}
+ \sqrt{3} {\rm e}^{-3\eta^2(1-p)t} \right] \nonumber \\ & & \times \left\{ 
\left[ \theta_3({\rm e}^{-\sqrt{3}t}) \theta_3({\rm e}^{-t/\sqrt{3}})
-1-\frac{\pi}{t} \right] +
\left[ \theta_2({\rm e}^{-\sqrt{3}t}) \theta_2({\rm e}^{-t/\sqrt{3}})
-\frac{\pi}{t} \right] \right\} . \label{J}   
\end{eqnarray}  
The first term on the rhs of (\ref{EV}) corresponds to the Coulomb energy
due to the non-neutrality of plate's entities.
\end{widetext}


\begin{thebibliography}{10}
\bibitem{rque51}
 {Within Poisson-Boltzmann theory, surfaces with surface
potentials of the same sign may attract, while surfaces with opposite surface
charges may repel, at short enough distances \cite{Chan}. 
Yet, for surface charges of the same sign, Poisson-Boltzmann yields repulsion. 
This is the situation we investigate in this work.}
\end{thebibliography}
\end{document}